\documentclass{iopjournal}

\usepackage{amsmath} 
\usepackage{cite} 

\begin{document}

\articletype{Research article} 

\title{Density-gradient effect in high-harmonic generation in gases}

\author{Zolt\'an Filus$^{1,*}$\orcid{0000-0000-0000-0000}, T\'imea Gr\'osz$^1$\orcid{0000-0000-0000-0000}, Chinmoy Biswas$^1$\orcid{0000-0000-0000-0000}, L\'en\'ard Guly\'as Oldal$^1$\orcid{0000-0000-0000-0000}, Tam\'as Bartyik$^1$\orcid{0000-0000-0000-0000}, Barnab\'as Gilicze$^1$\orcid{0000-0000-0000-0000}, Subhendu Kahaly$^{1,2}$\orcid{0000-0000-0000-0000}, Katalin Varj\'u$^{1,3}$\orcid{0000-0000-0000-0000} and Bal\'azs Major$^{1,3,*}$\orcid{0000-0000-0000-0000}}

\affil{$^1$ELI ALPS, The Extreme Light Infrastructure ERIC, Wolfgang Sandner u. 3, 6728 Szeged, Hungary}

\affil{$^2$Institute of Physics, University of Szeged, D\'om t\'er 9, 6720 Szeged, Hungary}

\affil{$^3$Department of Optics and Quantum Electronics, University of Szeged, D\'om t\'er 9, 6720 Szeged, Hungary}

\affil{$^*$Authors to whom any correspondence should be addressed.}

\email{Zoltan.Filus@eli-alps.hu}
\email{Balazs.Major@eli-alps.hu}


\begin{abstract}
High-harmonic generation (HHG) in gaseous targets is the most widespread method to produce coherent extreme-ultraviolet (XUV) pulses with sub-femtosecond duration. However, this process has intrinsically low efficiency, and substantial research and development is devoted worldwide to increase the achievable photon flux through this highly nonlinear light--matter interaction process. In this work, we show the strong interplay of phase matching and absorption in gas-pressure gradients, substantially affecting macroscopic HHG efficiency. Through detailed experimental analysis and supporting numerical studies, we highlight their significance, particularly at the boundaries of the interaction volume. The concluded results have implications in the massively expanding application possibilities of HHG sources requiring high photon flux, for example in the semiconductor industry, in nanoscale imaging of biological and industrial samples, or in nonlinear optics in the XUV regime.    
\end{abstract}

\section{Introduction}
A key discovery that subsequently led to the birth of attosecond physics was the first observation of high-order harmonic generation (HHG) in gases \cite{Ferray1988JPB, McPherson1987JOSAB}. Since then, blooming fields of science have emerged \cite{Krausz2009RMP, Calegari2016JPB, Li2020NC,Borrego-Varillas2022RPP, DiPalo2024APLPhoton}, with far-reaching possibilities in chemistry \cite{Lepine2014NatPhoton, Nisoli2017CR}, novel electron microscopy techniques \cite{Nabben2023Nature, Hui2024SciAdv, Gaida2024NatPhoton}, or in quantum physics \cite{Cruz-Rodriguez2024NRP}. 
The first industrial applications are also appearing thanks to new imaging techniques enabled by these light sources \cite{Eschen2022LSA, Tanksalvala2022SciAdv, Liu2023Photonix}.  State-of-the-art HHG sources are currently capable of delivering coherent extreme-ultraviolet (XUV) and X-ray pulses, with photon energies extending from a few tens of eVs to the keV regime \cite{Popmintchev2012SCI}, pulse durations below 100~attoseconds \cite{Calegari2016JPB}, and repetition rates exceeding 10~MHz \cite{Hadrich2016JPB, Porat2018NPL}. 

Despite their substantial advances, these sources still suffer from a limitation, i.e. the low XUV photon flux, at least compared to other coherent light sources operating in these wavelength regimes (for example, (X-ray) free-electron lasers \cite{Franz2024NatPhoton}). This is due to the inherently low fundamental-to-harmonic conversion efficiency, typically around $0.001\%$ when using lower-repetition-rate ($<1\,\mathrm{kHz}$), near-infrared (NIR, $\sim 1\,\text{\textmu m}$ wavelength) laser drivers and argon (Ar) gas as the target \cite{Rudawski2013RSI, Makos2020SR, Fabris2015NPhoton, Timmers2016Optica}. Conversion efficiency is known to decrease with higher repetition rates \cite{Saule2019NatComm, Ye2022UFS}, being closer to $10^{-7} - 10^{-8}$ in the hundreds of kHz  \cite{Rothhardt2014NJP, Hadrich2014NatPhoton, Ye2022UFS, Manschwetus2026SPIEProc} and tens of MHz range \cite{Saule2019NatComm}. A key factor identified for this trend is the formation of a steady-state plasma in the focal region of the laser beam, which must be considered to mitigate the reduction in conversion efficiency associated with high-repetition-rate operation \cite{Porat2018NPL}. Irrespective of the laser repetition rate, the low conversion efficiency can be improved by approximately one order of magnitude --- or somewhat beyond --- when using lower ionization rate noble gases, like xenon (Xe) \cite{Hadrich2015LSA, Porat2018NPL, Vernaleken2011OL, Ozawa2015OE} or krypton (Kr) \cite{Manschwetus2026SPIEProc}, or when using a shorter-wavelength driver system, for example second harmonic of an NIR pulse \cite{Wang2015NC, Klas2016Optica, Jurkovicova2024CommPhys}. These parameter configurations, however, are generally less widely available in laboratories. 

Optimization of the harmonic conversion efficiency is therefore critical to fully exploit the capabilities of HHG-based XUV sources. Such an optimization depends upon tuning the phase matching and reabsorption effects during the HHG process, which requires control of the macroscopic generation conditions. Attosecond science has an extensive literature, including papers that describe in detail the macroscopic generation process \cite{Gaarde2008JPB, Hareli2020JPB, Weissenbilder2022NRP, Appi2023OE}. One of these is a seminal theoretical work published by Constant et al. \cite{Constant1999PRL} characterizing the behavior of the XUV flux generated in media with a uniform density profile along the propagation axis. Although this theory has widely proved its applicability since its publication in 1999, it relies on a substantial simplification: the assumption of a flat-top gas density distribution in the interaction volume. This is an idealized approximation of the experimentally achievable density profiles, and --- more importantly --- does not consider potential effects that can arise due to the presence of pressure variations in the generation volume. To account for these, the related theory requires further extensions, e.g. to consider the finite density gradients closing the entrance and exit boundaries of the generation medium. A recent theoretical work \cite{Major2021JPB}, using the first-order extension of the model developed by Constant et al. \cite{Constant1999PRL}, studied exactly this situation, and predicted the dependence of the generated XUV flux on the density profile. Some aspects of this theoretical result appeared in the experimental observation by Ansari et al. \cite{Ansari2022JPB}, who reported on HHG from a neon-filled cell. A detailed analysis, however, was missing. 

In this work, we present a comprehensive theoretical study and the experimental verification of the critical role of a density-gradient effect on HHG flux, and demonstrate the importance of having a close to flat-top pressure profile for the HHG target. We provide an extended macroscopic generation model for HHG in absorbing gases, emphasizing that steep density gradients (approximating the idealistic assumption of a flat-top target by Constant et al. \cite{Constant1999PRL}) maximize photon flux for applications. With a careful design of the target gas cell, we achieve appropriate control over the experimentally obtainable density gradients, and confirm the validity of the extended model, highlighting the importance of the density gradient to boost the number of XUV photons available.

\section{\label{sec:methods}Materials and methods}

\subsection{\label{subsec:experiment}Experimental}

The experiments were carried out on the HR Gas beamline (High Repetition rate laser system-based gas high-order harmonic generation beamline for Gas targets) \cite{Ye2020JPB, Ye2022UFS} of the Extreme Light Infrastructure ERIC Attosecond Light Pulse Source (ELI ALPS) facility \cite{Kuehn2017JPB, Shirozhan2024UFS}. The HR Gas beamline can currently be driven by the 100-kHz repetition-rate HR1 \cite{Hadrich2022OL}, or the tuneable repetition-rate (10~Hz~--~10~kHz) HR Alignment lasers \cite{Gilicze2025HPL}. The results presented herein were obtained with the latter system, which is based on a Yb:KGW frontend with one nonlinear compression stage (multipass cell), and was available to provide ultrashort laser pulses at a 10~kHz repetition rate during these experiments.

\begin{figure}
\includegraphics[width=0.95\textwidth]{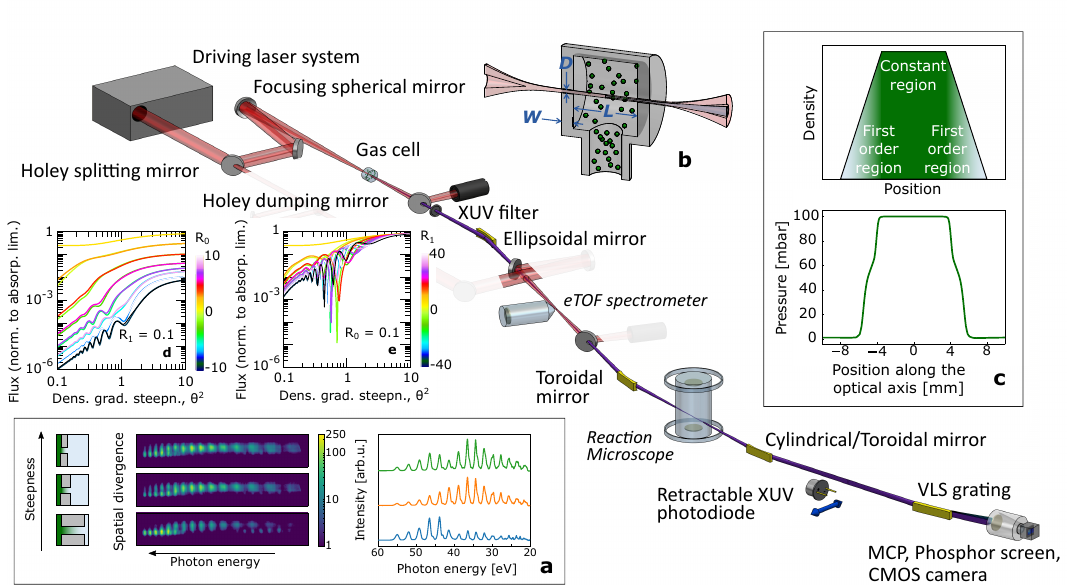}
\caption{\label{fig:scheme} Schematic optical layout of the relevant part of the HR Gas beamline used for the experiments. The functionality and type of the components are mentioned in the corresponding labels next to the graphics. Inset (a) shows the schematics of three significantly different apertures leading to different density gradients, along with the related spectrometer camera images and the calibrated XUV spectra. Inset (b) shows the schematics of the gas cells indicating the aperture diameter ($D$), wall thickness ($W$), and cell length ($L$) parameters, which are the control knobs for density gradients. Inset (c) is a graph pair demonstrating the density profiles of the gas medium along the laser propagation axis as assumed in the corresponding theoretical model (top) and as obtained from fluid dynamics calculations for $D$~=~0.5~mm and $W$~=~1.5~mm. Insets (d) and (e) show the theoretically predicted XUV flux as a function of density gradient steepness $\Theta^2$ for different phase matching conditions quantified by $R_0$ and $R_1$ (see Section \ref{subsec:theory}) for details of parameters and theory.}
\end{figure}

The compressed output of the HR Alignment laser ($\lambda$~=~1030~nm, 0.68~mJ, 34~fs) was employed to deliver the fundamental infrared (IR) driving field to the HR Gas beamline in its standard Mach–Zehnder interferometer arrangement, designed for attosecond pump-probe experiments (Figure~\ref{fig:scheme}). The setup was optimized to maximize the intensity of the fundamental beam in the XUV generating arm to 90\% of the input power. For the present experiment, simplifying the beamline to a single XUV generating arm, we used the standard configuration where the annular-shaped split beam was focused onto the gas target by a concave spherical mirror with a focal length of $f$~=~1500~mm. This loose focusing geometry yielded a focal spot size of 52~$\mu$m (full width at half maximum), with an estimated peak intensity of $\sim1.3\times10^{14}$~W/cm\textsuperscript{2}, accounting for imperfections of the focused beam profile, temporal contrast and reflection losses, along with the 90:10 reflected-to-transmitted splitting ratio for the input beam (see details in Supplementary material, Section III.C). The peak intensity estimation also fits with the highest harmonic order observed, based on the cut-off law of HHG \cite{Heyl2017JPB}.

This configuration led to XUV radiation generated in the $\sim25-50\,\mathrm{eV}$ photon energy range (see representative XUV spectra in Figure~\ref{fig:scheme}a, which show the known divergence variation of different harmonic orders in case of a gas targets placed before the laser focus \cite{Wikmark2019PNAS}). The generation medium was Ar gas in an in-house developed gas cell (see Figure~\ref{fig:scheme}b), designed to approach a trapezoidal density profile along the propagation axis (Figure~\ref{fig:scheme}c) to achieve the necessary control over the density gradients in the gas-filled volume. The key geometric parameters to achieve this control were the diameter $D$ of the orifice and aperture wall thickness $W$ (see Figure~\ref{fig:scheme}b for schematic gas cell geometry, and the detailed cell design in Supplementary material, Section III.B). As shown in Figure~\ref{fig:scheme}c, the entrance and exit regions of the medium can be approximated with linear density profiles, representing constant density gradients formed by the windowless apertures of the gas cell. The close-to-trapezoidal shape of the gas distribution was necessary for direct comparison with simulation results (see Section \ref{subsec:theory}).

Downstream of the target cell, the fundamental IR beam exited the HHG gas volume in a divergent annular form, which facilitated its separation from the generated high-harmonic (HH) field, as well as its dumping by a holey plane mirror equipped with an air-side beam dump and an anti-reflection-coated vacuum viewport. Following the dumping stage, optional metallic filters can be inserted in the beam path for further suppression of the residual IR beam, additional spectral tailoring of the generated XUV frequency comb, and compensating the attochirp if the eventual experiment so requires. The HH radiation then propagated through consecutive refocusing/reimaging stages, as the HR Gas beamline supports two experimental target areas, which were not utilized in the present experiments. In the last section of the HR Gas beamline, the HH radiation was directed to an XUV diagnostic section, where the pulse energy and spectral characteristics were monitored using a windowless aluminum oxide XUV photodiode (NIST~40790C) with calibrated spectral sensitivity, and an XUV spectrometer. The XUV spectrometer consists of a variable line spacing (VLS) spherical grating (HITACHI 001-0437, 1200~gr/mm) and a microchannel plate (MCP) detector system (Photek Ltd. VID140/P43/FOI) equipped with a high-speed CMOS camera (IDS UI-3060CP). Furthermore, the polarization state of XUV radiation can also be measured in this subsection of the beamline with a double-stage polarimeter \cite{GOL2025JPP}.

In this experimental campaign, various cell configurations were tested to explore the phase-matched parameter space at different backing pressures and cell positions within the confocal parameter range. The XUV photodiode served as a broadband calibration tool for the FFS camera pixel intensities, following camera background subtraction and Jacobian correction. The measured HH frequency comb distribution and the photodiode sensitivity were considered as weighting factors in the calibration process. Harmonic-wise XUV pulse energies and their fluctuations were extracted as the final evaluated experimental results (further details on data processing are provided in the Supplementary material, Section III.D).


\subsection{\label{subsec:theory}Theoretical}

A practical, one-dimensional theoretical model on optimizing HHG in absorbing gases was given by Constant et al. \cite{Constant1999PRL} in 1999, and the conditions they proposed for ideal phase matching have been experimentally validated \cite{Popmintchev2012SCI, Hadrich2014NatPhoton, Hadrich2016JPB, Heyl2017JPB, Weissenbilder2022NRP}. This model assumed an ideal, flat-top pressure profile for the generation medium, not considering potential density distributions, which are experimentally always present \cite{Comby2018OE, Hagmeister2022APB, Nagyilles2023PRAppl, Shlomo2025PRR}. Accordingly, the model has recently been extended to consider the non-flat-top shape of atomic density (e.g. triangular), and it has been proposed that the more it resembles the ideal profile, the more likely it is to reach the maximum flux \cite{Major2021JPB}. This "first-order" extension of the Constant model is obtained by calculation of the harmonic flux along the optical axis as
\begin{equation}\label{eq:integral}
	S^{(q)} \propto \left|  \int_{0}^{L_{\mathrm{med}}}  A^{(q)} \rho(z)
	\exp \left( \mathrm{i} \left[\Delta k^{(q)}(z) + \mathrm{i} \kappa^{(q)}(z)\right]
	\left[L_{\mathrm{med}}-z\right] \right) \mathrm{d}z
	\right|^{2} \,,
\end{equation}
where $A^{(q)}$ is the strength of the generated field (amplitude of the single-atom response), $\rho(z)$ is the number density of atoms, $\Delta k^{(q)}(z)$ and $\kappa^{(q)}(z)$ are the phase mismatch and absorption, respectively, while $L_{\mathrm{med}}$ is the total length (including density gradients and the constant pressure region) of the generation medium. Superscript $(q)$ indicates dependence on harmonic order $q$. For the extension of the model, the treatment introduced in Ref. \cite{Constant1999PRL} is followed, extending it with $\Delta k^{(q)}(z)$ and $\kappa^{(q)}(z)$ dependence to account for pressure gradient effects. Assuming that $\rho(z)$, $\Delta k^{(q)}(z)$ and $\kappa^{(q)}(z)$ are either constant or change linearly with distance $z$ (like in the case of a trapezoidal pressure profile \cite{Major2021JPB}), it can be shown that the phase matching and absorption conditions can be parametrized by two dimensionless variables
\begin{equation}\label{eq:Rn}
    R^{(q)}_{j} = \Delta k^{(q)}_{j} / \kappa^{(q)}_{j}
\end{equation}
where $j=0$ or $1$, according to the coefficients of the first-order Taylor-like expansion for $\Delta k^{(q)}(z) = \Delta k^{(q)}_{0} + \Delta k^{(q)}_{1}\cdot z$ and $\kappa^{(q)}(z) = \kappa^{(q)}_{0} + \kappa^{(q)}_{1}\cdot z$ \cite{Major2021JPB}. For some specific cases even analytical solutions can be derived \cite{Major2021JPB}.

The $R_1$ and $R_0$ parameters in Eq. (\ref{eq:Rn}) describe phase matching conditions in the density gradient and constant pressure regions, respectively. Explicitly, $R_0 = \Delta k_0 / \kappa_0$ describes the ratio of phase mismatch $\Delta k_0$ and absorption $\kappa_0$ in the constant density region, while $R_1 = \Delta k_1 / \kappa_1$ is a similar ratio but for gradient quantities in the linearly-varying density region. Accordingly, quantities indexed by 0 and 1, e.g. $\Delta k^{(q)}_{0}$ and $\Delta k^{(q)}_{1}$, are different physical quantities with different dimensions, since $\Delta k^{(q)}_{0}$ is the phase mismatch in the constant pressure region, while $\Delta k^{(q)}_{1}$ is the rate of change of phase mismatch along the propagation axis in the density gradient region. Similarly, $\kappa^{(q)}_{0}$ is absorption in the constant pressure region (units of $1/\text{m}$), while $\kappa^{(q)}_{1}$ is the rate of change of absorption in the density gradient region (units of $1/\text{m}^2$). To simply follow the effect of these parameters on HHG, a lower value of $|R_{0,1}|$ means better phase matching (perfect phase matching is achieved with $R_0 = R_1 = 0$, see examples in Figure \ref{fig:scheme}e and d, along with details in Supplementary material Section I).

In this work, we numerically evaluated the integral in Equation~(\ref{eq:integral}) assuming a trapezoidal density profile (see Figure~\ref{fig:scheme}c) of the form
\begin{equation}
    \rho(z) = 
    \begin{cases}
        \rho_1 z &  \text{ if } 0 \leq z < 1.25 \cdot L_{10/90}, \\
        \rho_0  & \text{ if }  1.25 \cdot L_{10/90} \leq z \leq  1.25 \cdot L_{10/90} + L , \\
        \rho_0 - \rho_1 z & \text{ if } 1.25 \cdot L_{10/90} + L < z \leq L_{\mathrm{med}}
    \end{cases} 
\end{equation}
with a total medium length of $L_{\mathrm{med}} = L + 2 \cdot  1.25 \cdot L_{10/90}$, $L$ being the region with constant pressure, $z$ measured from the start of the trapezoidal pressure distribution, and $\rho_0$ and $\rho_1$ being the density and density gradients (with units of $1/\text{m}^3$ and $1/\text{m}^4$), respectively (see details in the Supplementary material, Section I). The density gradient is described by the transition length parameter ($L_{10/90}$), defined as the 10\% to 90\% transition length of the density profile variation, analogous to electronic signal rise/fall times measured between 10\% and 90\% of the voltage difference during signal transitions. We note here that defining the transition length with other values than 10\% and 90\% does not change the conclusions of this work, and values were chosen because of the above practicality.

To connect the $R_0$ and $R_1$ parameters to more widely used and physically related phase matching contributions \cite{Heyl2017JPB}, it can be shown that the phase mismatch in most experimental cases can be expanded to a constant and a density-proportional term \cite{Major2021JPB}. This leads to the relations:
\begin{equation}\label{R1}
    R_{1} = \frac{2(\gamma_n + \gamma_p)}{\sigma}
\end{equation}
and
\begin{equation}\label{R0}
    R_{0} = R_{1} + \frac{\Delta k_{\mathrm{const}}}{\kappa_{0}}\,,
\end{equation}
where the $\gamma_n$ and $\gamma_p$ are proportionality (with density) constants of neutral- and plasma-dispersion related phase mismatch \cite{Major2021JPB}, $\Delta k_{\mathrm{const}}$ is the phase mismatch term independent of pressure (geometrical and dipole phase related terms \cite{Heyl2017JPB}), and $\sigma$ is the photoionization cross-section. Superscript $(q)$, indicating the photon energy/harmonic order dependence, has been omitted here for all terms for conciseness. As has been shown, keeping $R_{j}$ values low is the key for phase-matched high-harmonic generation \cite{Major2021JPB}. The steepness of the density gradient can also be quantified using a dimensionless quantity
\begin{equation}
    \Theta^2 = \kappa_{1} / \kappa_{0}^2\,,
\end{equation}
which allows one to analyze the effect of this property generally, and simply scale results to the specific experimental conditions \cite{Major2021JPB}. Accordingly, throughout this work, we will use $\Theta^2$ to quantify the steepness of the linear density gradient, since theory shows that this value determines the effects of linear density gradients on HH flux, regardless of what technical parameter of the gas target is used to alter the gradient steepness (for example, the cell aperture size $D$, wall thickness $W$, or gas pressure $p_{IN}$, see in Section \ref{sec:results}). Detailed description of the phase matching and absorption models, and calculation of the values of $R_{j}$ and $\Theta^2$ for our experimental conditions along with the simulations of achievable photon flux using Equation~(\ref{eq:integral}) are given in the Supplementary material, Section I.
We add here --- as it served as the main motivation of the present work --- that remarkably, the theoretical analysis shows that steeper gradients (larger value of $\Theta^2$) lead to higher flux irrespective of phase matching conditions represented by parameters $R_1$ and $R_0$ in Figure~\ref{fig:scheme}d and e.

To match with the theoretical study of HH flux and in order to design appropriate gas targets for our analysis, gas distribution calculations were carried out with the Simcenter FloEFD (Siemens AG) computational fluid dynamics package. All simulations were conducted with a fixed cell length $L$ of 3~mm (as verified later, longer cell lengths made no relevant difference in the achieved density gradients), focusing on the influence of varying geometric parameters, starting from a baseline configuration comprising a backing pressure of 100~mbar, an orifice diameter ($D$) of 700~$\mu$m, and a wall thickness ($W$) of 0.5~mm. Details of the fluid dynamics simulations are given in Supplementary material Section III.B.2.

\section{\label{sec:results}Results}

\begin{figure}
\includegraphics[width=0.9\textwidth]{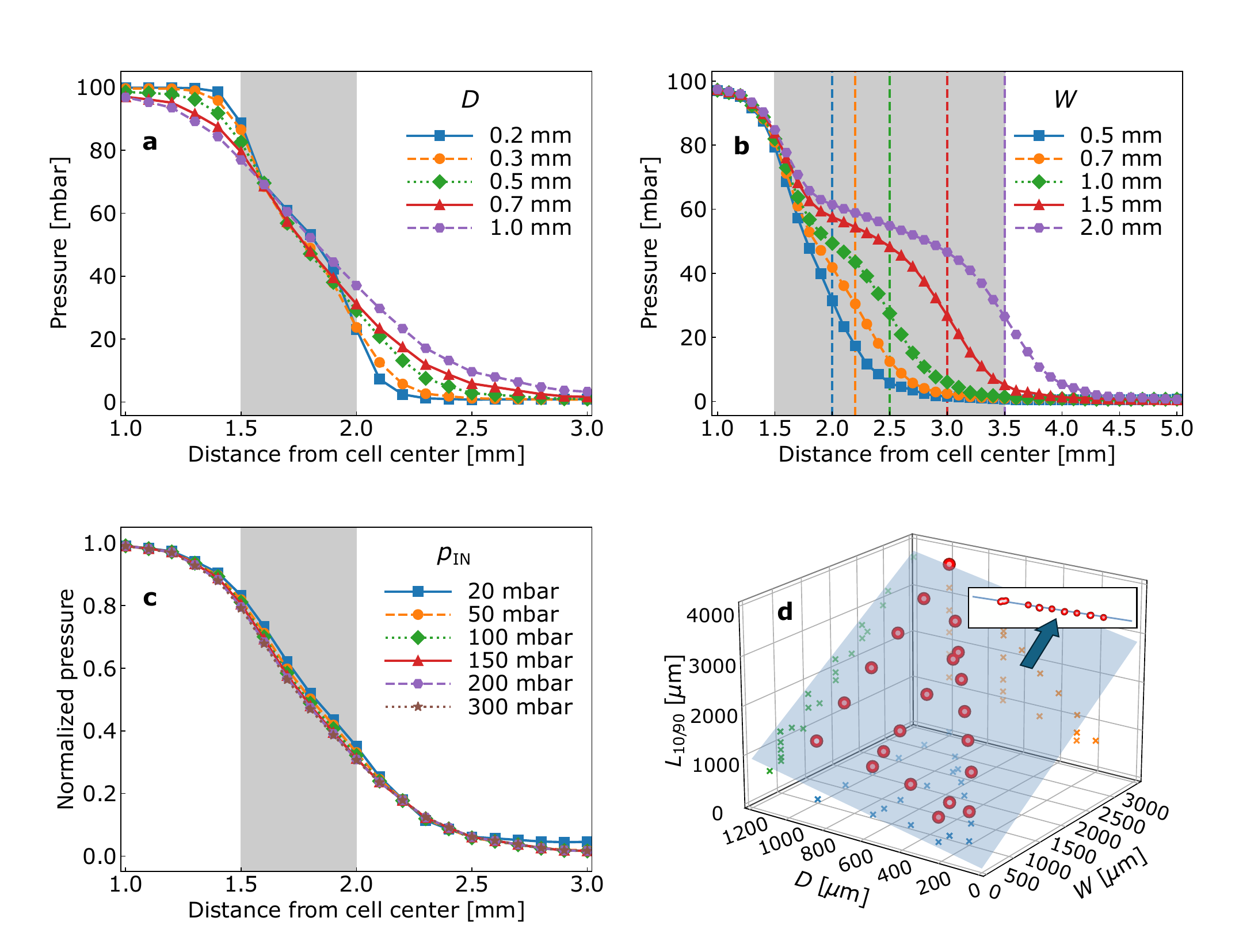}
\caption{\label{fig:gradient} Simulated pressure profiles along the propagation axis of the laser beam around the aperture regions of the HHG cell (a-c) based on fluid dynamics modeling. The horizontal axes in subplots (a-c) represent distances measured from the cell center. The graphs demonstrate the dependency of the pressure profile on the orifice diameter $D$ (a), the aperture wall thickness $W$ (b), and the backing pressure $p_{\mathrm{IN}}$ (c). The gray region of subplots a to c symbolizes the aperture wall; the white regions on the left and right sides represent the cell interior and the external vacuum space around the cell, respectively. Subgraph (d) shows the combined orifice diameter and wall thickness dependency of the transition length $L_{10/90}$ (length of 10\% to 90\% pressure increase, see text) and the best fit plane serving the empirical formula for the transition length in Equation~(\ref{eq:L1090}). The inset demonstrates the goodness of the fit of the plane on the simulated points by an in-plane viewpoint.}
\end{figure}

The primary objective of our research was to study the effect of pressure gradients on the achievable photon flux through investigating the efficiency of the HHG process using various gaseous target media characterized by different cavity pressures and pressure gradients at the boundary of the gas cell and vacuum environment. To control the pressure gradients, we manufactured a series of gas cells with different aperture geometries, which define the achievable gradient (the value of $\Theta^2$). We systematically analyzed the efficiency results by comparing them against each other and against the corresponding theoretical model.

The actual gas target technically consisted of a multi-cell system mounted on a 5-axis translational-rotational stage for precise alignment and positioning. The cell length was set to 8~mm, roughly corresponding to the constant-density section of the trapezoidal profile (see  Figure~\ref{fig:scheme}b and c). This length was chosen to achieve an absorption limited flux in the central part of the generation medium, consistent with the criterion established by Constant et al. \cite{Constant1999PRL}, according to whom the constant-density medium length $L$ should exceed the absorption length $L_{\mathrm{abs}}$ by a factor of 3 ($L > 3L_{\mathrm{abs}}$). This condition is satisfied here, as the absorption length in argon ranges from several hundred microns to a few millimeters within the XUV wavelength and pressure ranges relevant to our experiments \cite{elettra_web} (see absorption data in the Supplementary material, Section I.B.5). The cells were specifically designed to allow variability in the input/output aperture walls. For this purpose, a set of aperture walls was manufactured with systematically varied aperture diameters and wall thicknesses, covering approximately a 12-fold range in the 10-to-90\% transition length $L_{10/90}$ of the density profile variation. The cells were backed with Ar gas in a range from 60~mbar to 140~mbar, allowing exploration of a $\Theta^2$ gradient range spanning roughly a factor of 28 between the steepest and shallowest pressure gradients. The well-controlled, linear density gradients were necessary as numerical analysis showed that the photon flux of HHG sources heavily depends on the gradient steepness (see Figure~\ref{fig:scheme}d and e).

Figure~\ref{fig:gradient} shows density profiles derived from fluid dynamics simulations of the gas cell volume and its surroundings (see details in the last paragraphs of Section~\ref{subsec:theory} and Supplementary material, Section III.B). Due to the mirror symmetry in the density profiles along the laser propagation direction, the corresponding graphs (Figure~\ref{fig:gradient}a-c) display only the segments where gas density distributions change with control parameters.   The transition length was efficiently controlled by adjusting the parameters highlighted in Figure~\ref{fig:scheme}b: diameter $D$ of the orifice (Figure~\ref{fig:gradient}a) and aperture wall thickness $W$ (Figure~\ref{fig:gradient}b), while it remained nearly independent of the backing pressure $p_{\mathrm{IN}}$ (Figure~\ref{fig:gradient}c) and cell length $L$ (not shown). The simulations were initiated from the baseline configuration ($p_{\mathrm{IN}}=100\,\mathrm{mbar}$, $D=700\,\text{\textmu m}$, and $W=0.5\,\mathrm{mm}$), with only the parameter of interest varied while all other parameters were kept unchanged. This approach allowed the individual influence of each parameter on the resulting density profile and transition length to be evaluated independently. For the three-dimensional representation of Fig.~\ref{fig:gradient}d, additional simulations were performed for the specific $D-W$ combinations corresponding to the gas-cell aperture configurations manufactured and tested experimentally. Finally, additional simulations, not shown here, were performed to verify the invariance of the resulting density gradients with respect to the diameter of the gas support tube, the choice of test gas, and the cell length.

Within the experimentally relevant and technologically feasible (limited by the $W<3D$ relationship for mechanically fabricated apertures) range of these geometric parameters, the transition length follows the empirical linear relationship
\begin{equation}\label{eq:L1090}
    L_{10/90} = 0.763 \cdot D + 0.955 \cdot W \,,
\end{equation}
where the units of all quantities are lengths in microns (Figure~\ref{fig:gradient}d). As shown in Figure \ref{fig:gradient}d and Equation~(\ref{eq:L1090}), there is a slightly stronger dependence of transition length on wall thickness $W$ compared to aperture diameter $D$.
The transition lengths were converted to the dimensionless parameter $\Theta^2$ for comparison with theory in the case of the different pressures and harmonic orders (see details in the Supplementary material, Section I.B.7). During the experimental campaign, a wide range of aperture walls were tested, covering transition lengths from $L_{10/90}=0.5\,\mathrm{mm}$ up to $L_{10/90}=6.0\,\mathrm{mm}$, spanning well beyond an order of magnitude of $\Theta^2$ values ($\sim0.1$ to $\sim10$, the exact range depending on the backing pressure and photon energy (harmonic order)).

\begin{figure}[htb!]
\includegraphics[width=0.95\textwidth]{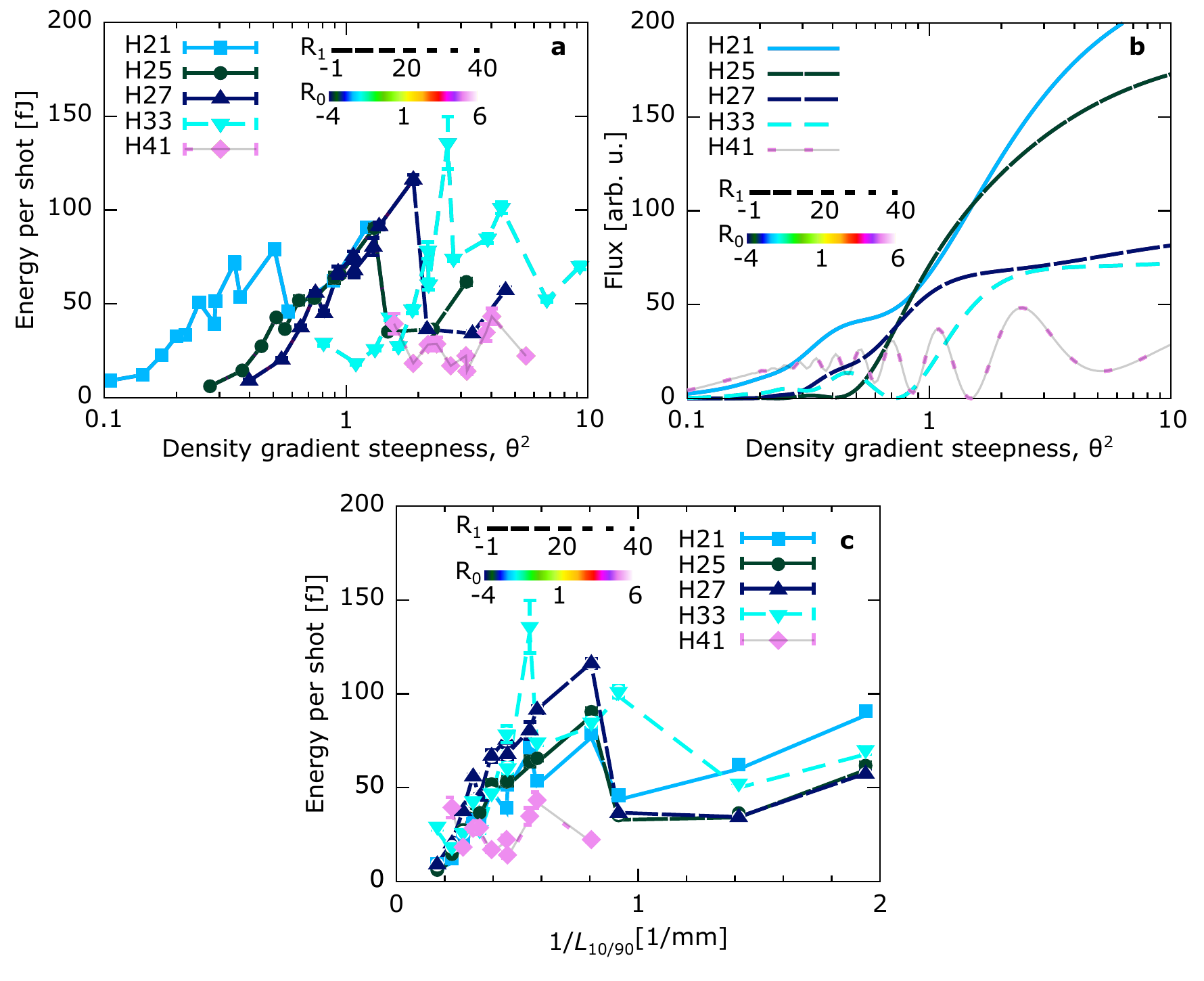}
\caption{\label{fig:Hdep} (a) Measured and (b) simulated flux at a fixed, $90\,\mathrm{mbar}$ gas cell pressure as a function of pressure gradient steepness $\Theta^2$ for different harmonic orders (in the range H21-H41, $25.3-49.4\,\mathrm{eV}$). The value of phase matching parameter $R_0$ is coded by the color of the curves according to the colorbar, while the other parameter $R_1$ is given by the dash-length in each curve. The specific values are $R_0 = -2.42, -3.64, -3.90, -1.88,  5.07$ and $R_1 = -0.68, 1.53, 4.12, 17.54, 39.06$ for H21, H25, H27, H33, H41, respectively. Both the measured and simulated results show the overall increase of harmonic flux with pressure with different amounts of oscillations for different harmonic orders. (c) Measured flux with the same parameters as in (a) but plotted as a function of reciprocal transition length $1/L_{10/90}$ to represent direct connection to this experimental parameter.}
\end{figure}

\begin{figure}[htb!]
\includegraphics[width=0.95\textwidth]{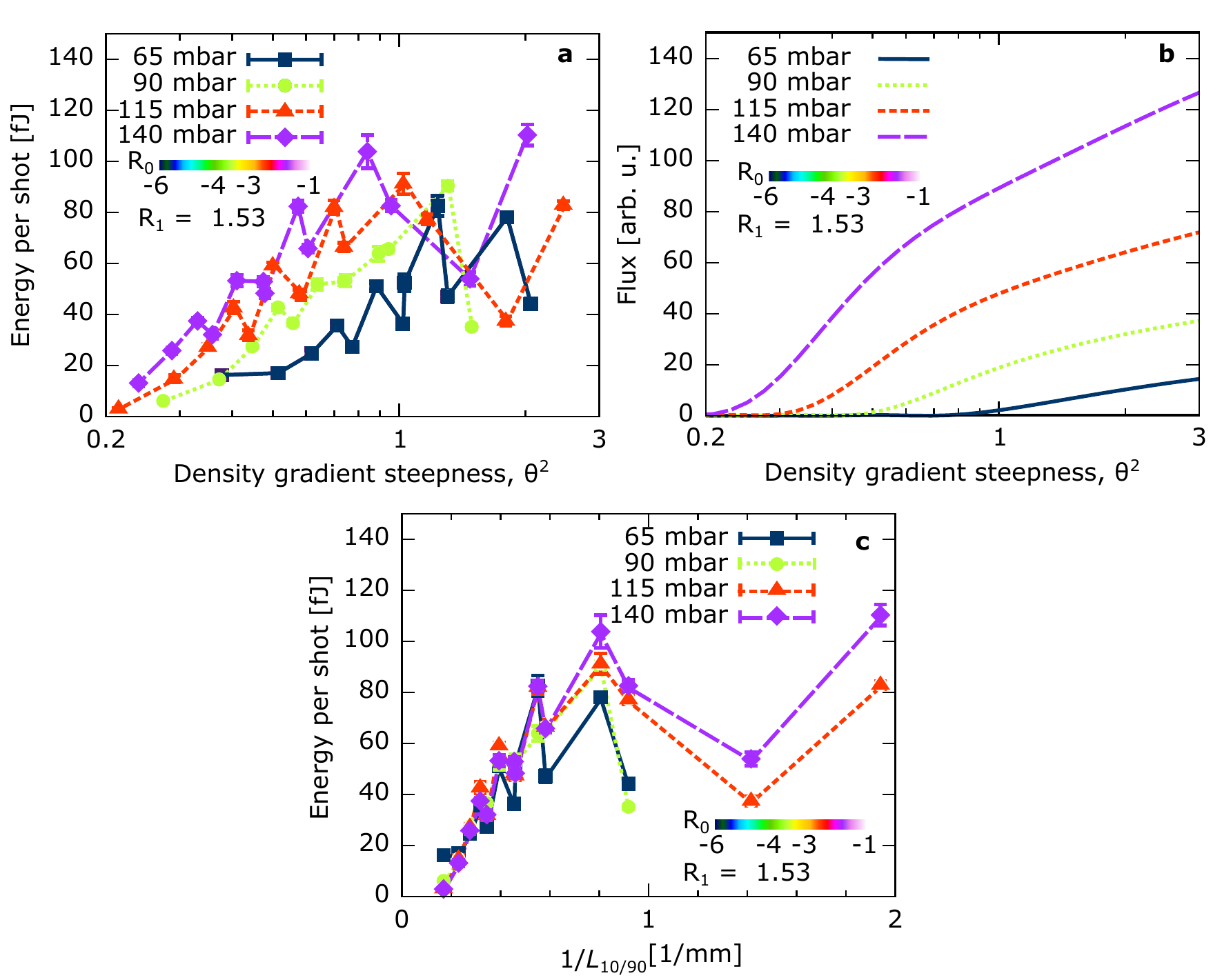}
\caption{\label{fig:pdep} (a) Measured and (b) simulated flux of harmonic order H25 ($30\,\mathrm{eV}$) as a function of pressure gradient steepness $\Theta^2$ for different gas pressures in the gas cell. The value of phase matching parameter $R_0$ is coded by the color of the curves according to the colorbar, while the other phase matching parameter is fixed to $R_1 = 1.52$, as it is independent of pressure (see Supplementary material, Section I.B.6). Specifically, $R_0 = -5.63, -3.64, -2.52, -1.80$ for $65\,\mathrm{mbar}$, $90\,\mathrm{mbar}$, $115\,\mathrm{mbar}$, $140\,\mathrm{mbar}$, respectively. (c) Measured flux with the same parameters as in (a) but plotted as a function of reciprocal transition length $1/L_{10/90}$ to represent direct connection to this experimental parameter. The measured and simulated results show the same increase of harmonic flux with pressure and its gradient steepness, independent of phase matching conditions, which are different for different pressures. }
\end{figure}

Finally, the full set of recorded XUV pulse energies was analyzed for each harmonic order as a function of the density gradient at the boundaries of the generation medium. The results are summarized in Figure~\ref{fig:Hdep} and Figure~\ref{fig:pdep}, which plot the steepness parameter $\Theta^2$ on the horizontal axis and XUV pulse energies on the vertical axis. The error bars represent the standard deviation of pulse energy fluctuations obtained from repetitive measurements with the flat-field photon spectrometer. Further details on data processing can be found in Supplementary material, Section III.D.

Figure~\ref{fig:Hdep} illustrates the harmonic order dependence of the XUV flux at a backing pressure of 90~mbar, comparing the experimental data (Figure~\ref{fig:Hdep}a) and the simulated results (Figure~\ref{fig:Hdep}b). In a similar scheme, Figure~\ref{fig:pdep} presents backing pressure dependent plots of the XUV flux for the 25\textsuperscript{th} harmonic order (H25, $\sim30\,\mathrm{eV}$), showing both measured and simulated data. Although plotting the flux as a function $\Theta^2$ provides better comparability with simulations (also from a practical perspective, as curves are better separated in the parameter space), we additionally plot the same information as a function of $1/L_{10/90}$ in Figure \ref{fig:Hdep}(c) and Figure \ref{fig:pdep}(c). This provides a possibility to directly see the results as a function an implicit experimental parameter. Conclusions are not affected by what is chosen as the horizontal axis. 

The measurement data points in plots these plots (Figure~\ref{fig:Hdep}(a) and (d, Figure~\ref{fig:pdep}(a) and (d)) were obtained by measuring HHG flux with various cell apertures and wall thicknesses. We note here, that the maximum energy per shot in a single harmonic peak --- having a maximum of around $150\,\mathrm{fJ}$ in Figure~\ref{fig:Hdep} and Figure~\ref{fig:pdep} --- might seem low, but considering that we have 15 harmonic orders present in the plateau (see Figure~\ref{fig:scheme}(a)), and we have $~\sim19\%$ transmission from the HHG source to the target area (where XUV energy is measured) \cite{Ye2022UFS}, the generated energy per shot at source totals in $12\,\mathrm{pJ}$. This means $\sim6\cdot10^{-8}$ conversion efficiency with our effectively $0.196\,\mathrm{mJ}$ laser pulses (see details in Supplementary material, Section III.C), typical for broadband sources operating at $100\,\mathrm{kHz}$ repetition rate \cite{Rothhardt2014NJP, Hadrich2014NatPhoton, Ye2022UFS, Manschwetus2026SPIEProc}. 

\section{\label{sec:discussion}Discussion}

The gradient-dependent plots presented in this work demonstrate good agreement between experimental and theoretical results under identical HHG conditions shown in Figure~\ref{fig:Hdep}a and b for harmonic order dependence, and Figure~\ref{fig:pdep}a and b for pressure dependence. This latter --- the pressure dependence --- shows slightly lower variation in the experiment, but the trends of increasing flux with gradient steepness is clearly matching.  The modeled curves and the experimental observations follow and support each other in the entire, experimentally accessible steepness parameter range covering two orders of magnitude. The agreement between the experimental data and the theoretical curves confirms that our model catches the key aspects of the macroscopic HHG process and describes the underlying physics of density-gradient effects.

As is predicted by theory (see Figure~\ref{fig:scheme}d and e) and confirmed by our wide parameter range measurements (Figure~\ref{fig:Hdep} and Figure~\ref{fig:pdep}), irrespective of phase matching conditions (different values of $R_{0}$ and $R_{1}$, see specific values in captions of Figure \ref{fig:Hdep} and \ref{fig:pdep}), there is an increase in XUV photon flux with increasing density gradient steepness (increasing value of $\Theta^2$). This can be explained by two key factors. First, even if perfect phase matching is achieved in the constant pressure region of the interaction volume, it is detuned in the density gradient, and destructive interference builds up as the XUV radiation leaves the interaction area. Second, in a longer gradient, the field potentially can be absorbed more. A shorter propagation path in a density gradient at the end of the interaction volume provides more efficient phase matching and minimum absorption.  

The results show a strong harmonic order dependence on the achievable flux, amplifying the lower orders more in comparison to the higher ones.  This finding implies that gradient optimization can be selectively tuned to support specific spectral regions, offering extra freedom to control the HHG output according to the specific needs of an application. Furthermore, higher harmonic orders show a wider steepness dependence and less or even oscillating harmonic flux. These are primarily due to the smaller value of $|R_{0,1}|$ for lower-order harmonics in our case (see values in the caption of Figure \ref{fig:Hdep}). As predicted by theory (see Figure \ref{fig:scheme}d and e), in the case of bad phase matching conditions (larger values of $|R_{0,1}|$), the flux strongly oscillates as a function of gradient steepness, as in the case of H41 in Figure \ref{fig:Hdep}. This is because the phase mismatch $\Delta k$ becomes so large that in the variation range of the transition length $L_{10/90}$ the oscillations with medium length become visible.  Considering the pressure dependence (Figure \ref{fig:pdep}), one can see clear monotonic trends as a function of gradient, intensifying the fluxes associated with higher backing pressures more significantly than the lower ones. 

\begin{figure}[htb!]
\includegraphics[width=0.95\textwidth]{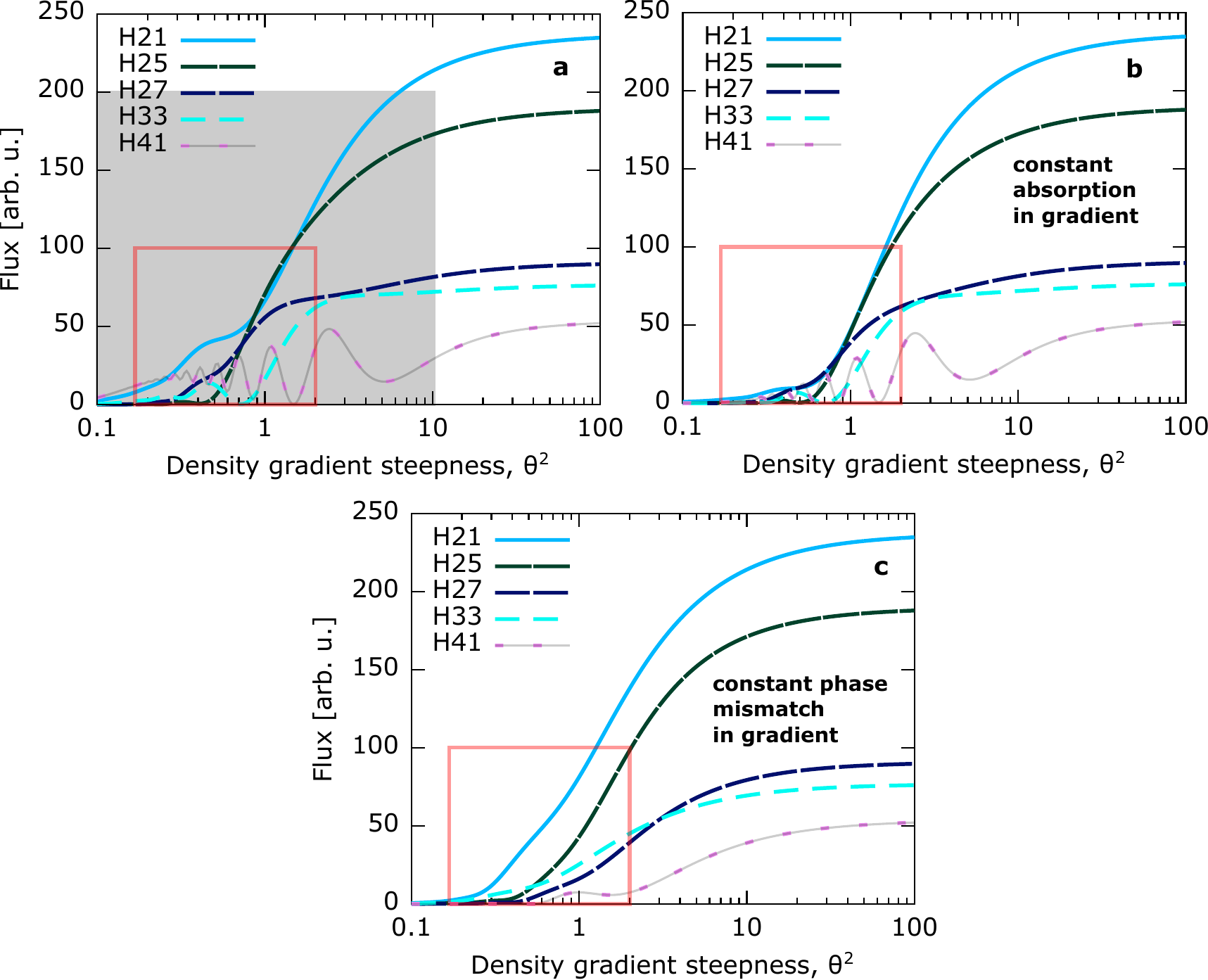}
\caption{\label{fig:Hdep_abs_pm} Simulated harmonic flux as a function of density gradient steepness with (a) the same parameters as in Figure \ref{fig:Hdep} but in a broader steepness range, (b) when the absorption is set to a constant value in the density gradient (equivalent to the absorption in the constant density area), and (c) when the phase mismatch is set to a constant value in the density gradient (equivalent to the phase mismatch in the constant density area). The gray background in (a) shows the axes ranges of Figure \ref{fig:Hdep}. The orange rectangle aims to highlight the gradient steepness range in which differences are biggest in between the three cases. }
\end{figure}

\begin{figure}[htb!]
\includegraphics[width=0.95\textwidth]{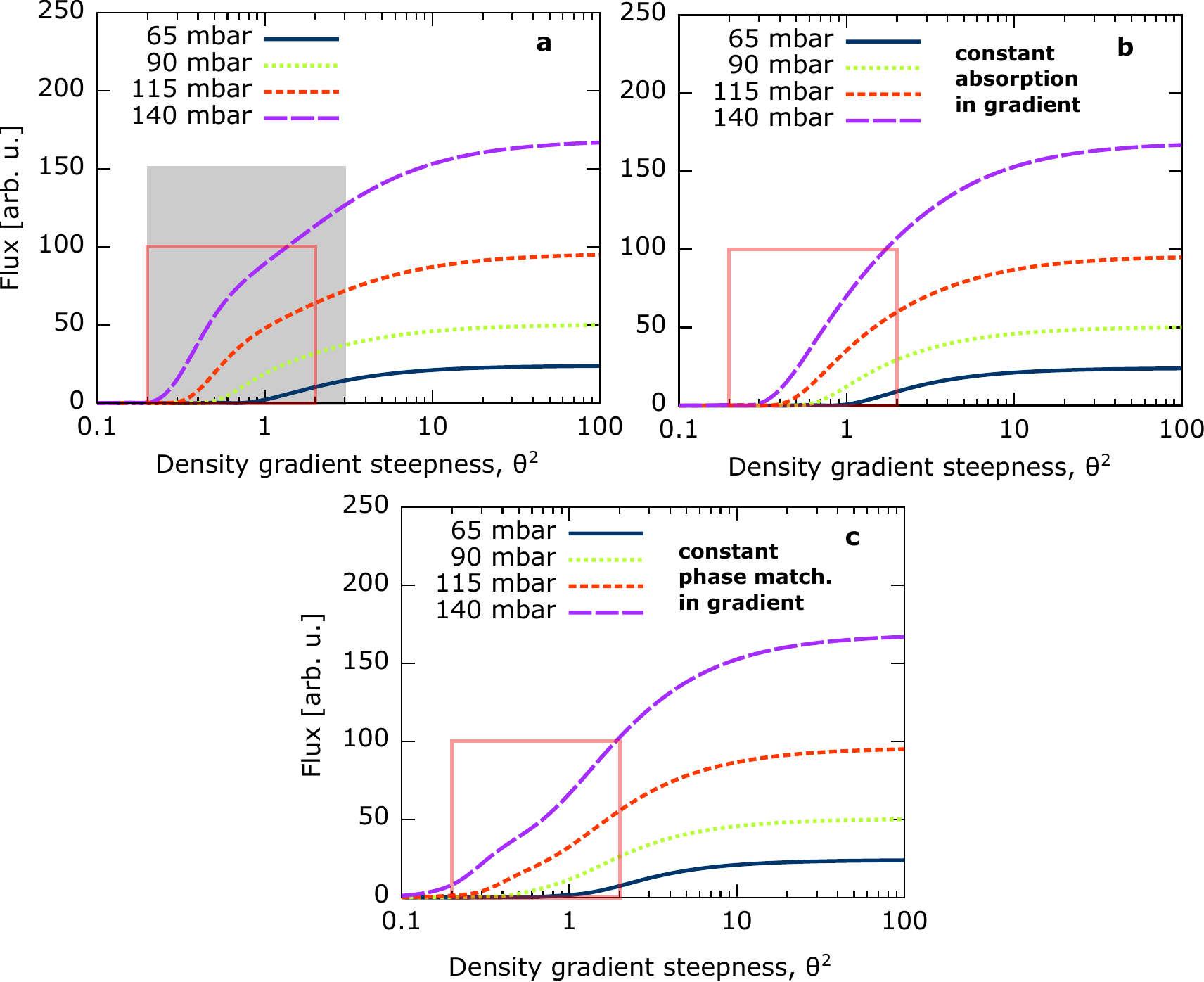}
\caption{\label{fig:pressdep_abs_pm} Simulated harmonic flux as a function of density gradient steepness with (a) the same parameters as in Figure \ref{fig:pdep} but in a broader steepness range, (b) when the absorption is set to a constant value in the density gradient (equivalent to the absorption in the constant density area), and (c) when the phase mismatch is set to a constant value in the density gradient (equivalent to the phase mismatch in the constant density area). The gray background in (a) shows the axes ranges of Figure \ref{fig:pdep}. The orange rectangle aims to highlight the gradient steepness range in which differences are biggest in between the three cases. }
\end{figure}

To strengthen our claim that both phase mismatch and absorption play a key role in the observed HH flux changes with gradient steepness --- and it is not simply due to reabsorption during the additional propagation in the gradient --- we carried out modified simulations presented in Figures \ref{fig:Hdep_abs_pm} and \ref{fig:pressdep_abs_pm}. 
Figure \ref{fig:Hdep_abs_pm}a shows the simulated HH flux dependence with gradient steepness for the same parameters as in Figure \ref{fig:Hdep}b, but in an extended gradient range (the gray background in Figure \ref{fig:Hdep_abs_pm}a is the plot range of Figure \ref{fig:Hdep}b). In Figure \ref{fig:Hdep_abs_pm}b, simulation results are shown where we artificially used a constant absorption value even in the area where the pressure gradients are present. In a similar manner, in Figure \ref{fig:Hdep_abs_pm}c, the phase mismatch term was kept constant in the whole interaction region (with a value equivalent to the one representing the constant density region). As can be seen in these plots (especially in the range highlighted by the orange rectangle), neglecting either the rate of change of absorption or the rate of change of phase mismatch in the gradients modifies the shape of the curves, and does modify it in a similar extent. This confirms that both absorption and phase mismatch changes are important to consider. Of course, in case of very steep gradients (right to the orange rectangle highlight of \ref{fig:Hdep_abs_pm} and  \ref{fig:pressdep_abs_pm}) there is no observable difference between subfigures a, b and c, as there is short propagation distance for phase mismatch or absorption to be relevant. In case of very shallow gradients (left to the orange rectangle highlight of \ref{fig:Hdep_abs_pm} and \ref{fig:pressdep_abs_pm}), the long propagation leads to low flux due to absorption and phase mismatch. We also add here that, furthermore, in both the harmonic order and pressure dependent graphs (Figures \ref{fig:Hdep_abs_pm}a and \ref{fig:pressdep_abs_pm}a), the larger range of gradients forecasts an additional boost of flux in the extreme theoretical steepness range. 

From the practical point of view, significant improvements in HHG efficiency can be achieved when smaller and thinner orifices, i.e. steeper density gradients (higher values of $\Theta^2$), are used. These results indicate that careful engineering of the gas cell geometry, particularly through minimizing the orifice diameter and thickness, can yield remarkable gains in XUV flux output. Based on the empirical formula (Equation~\ref{eq:L1090}), we expect that lower transition length can be achieved by smaller apertures. In case of low average-powered driving laser systems, thin metallic foil-based aperture walls with laser-drilled orifices are applied \cite{Cardin2018JPB}, which could potentially provide the ideal gas target in terms of steep density gradients. The HHG efficiency is improved by close to a factor of ten with reference to the frequently used aperture configurations ($D\approx0.8\,\mathrm{mm}$, $W\approx0.9\,\mathrm{mm}$, $L_{10/90}\approx1.5\,\mathrm{mm}$) of the HR Gas beamline. This is a promising path for maximizing photon flux, as shown also by our analysis of previous works in the literature, for which we forecast a similar flux increase by tailoring the gas target according to our proposal (see examples of potential flux increase in Supplementary material, Section II). At the same time, considering the boosting effect, it is particularly critical to deal with the reduction of this efficiency in cases where this idealized limiting case is not technically feasible, for example because of the need to apply robust aperture walls against high-average-power driving laser systems \cite{Filus2022RSI}.

We also add here that the above considerations also apply to semi-infinite gas cells \cite{Vismarra2024}, where effectively only the end aperture exists. This is because previous theoretical analysis showed that the density gradient at the exit aperture is what defines these effects primarily \cite{Major2021JPB}. Detailed analysis of this type of target, however, lies outside the scope of this work.  

\section{Conclusions}

In summary, it can be concluded that control over the density gradient of the HHG medium is particularly important for achieving high XUV flux in HHG. Such a comprehensive experimental verification has been, to the best of our knowledge, given for the first time.

Our conclusion is additionally that the optimum cell design for exploiting opportunities in gas medium tailoring is the one minimizing the transition length. We expect that the ultimate minimum transition length can be realized by laser-drilled foil-wall gas cells, which must be combined with precise pointing stabilization systems in the case of high-average-power lasers, to avoid aperture damage and thereby keep the orifice diameter at the minimum value. 

Next steps of this research involve the design and experimental validation of gas cells with laser-drilled aperture walls under high-average-power conditions, leading to the characterization of the mechanical and optical limits of cell-based HHG systems. Toward the next-generation designs, extending the work to more realistic and complex gas density profiles offers the possibility of further flux enhancement. Importantly, such precise control over the density gradient is an intrinsic feature of gas-cell-based targets and cannot be achieved in free-expanding gas jets. These developments may include formation of cell-sized periodic density profiles for further improving XUV flux through quasi-phase matching processes \cite{Ciriolo2022APLPhoton} --- meaning additional density gradients appearing in the interaction volume --- that can be combined with advanced laser shaping and focusing techniques.

%
%

\ack{We acknowledge the Extreme Light Infrastructure (ELI) for providing access to experimental facilities and support. This work is based on experiments performed at the HR Gas beamline at the ELI ALPS Facility. The ELI ALPS project (GINOP-2.3.6-15-2015-00001) is supported by the European Union and co-financed by the European Regional Development Fund. The work of B.M. was supported by the Bolyai J\'anos Research Scholarship of the Hungarian Academy of Sciences. }

\funding{European Union and the European Regional Development Fund (GINOP-2.3.6-15-2015-00001), Hungarian Academy of Sciences (Bolyai Research Scholarship (BO/00454/24))}

\roles{Z.F. and B.M. conceived the project. B.M. performed the theoretical analysis. Z.F. designed the gas cells. Z.F., T.G., C.B., L.G.O. and B.M. designed the experiments. Z.F., T.G., C.B., L.G.O., T.B. and B.G. performed the experiments. Z.F. and B.M. analysed the experimental data and wrote the manuscript, with input from the other authors.}

\data{The data that support the findings of this study are available from the corresponding authors upon reasonable request}

\suppdata{Supplementary material document with a more detailed description of methods.}

\bibliography{bibliography_pressgrad}
\bibliographystyle{iopart-num}

\end{document}